\documentclass[floats,floatfix,showpacs,amssymb,prd,twocolumn,superscriptaddress,nofootinbib,nolongbibliography,reprint]{revtex4-1}

\usepackage[utf8]{inputenc}
\usepackage{graphicx,amssymb,amsmath,amsthm,amsfonts,epsfig,epsf,environ,float}
\usepackage[usenames,dvipsnames,svgnames,table]{xcolor}
\usepackage{epstopdf}
\definecolor{darkred}{rgb}{0.5,0,0}
\usepackage{bm}
\usepackage{dcolumn}
\usepackage{latexsym}
\usepackage{rotating}
\usepackage{longtable}

\usepackage{enumerate}
\usepackage{tensor,multirow}
\usepackage{url}
\usepackage[linktocpage]{hyperref}

\makeatletter
\NewEnviron{nofleqn}[1]{\@fleqnfalse\begin{#1}\BODY\end{#1}}
\makeatother

\hypersetup{
	colorlinks,
	linkcolor={red!60!black},
	citecolor={green!50!black},
	urlcolor={blue!80!black}
}

\def\be{\begin{equation}}
\def\ee{\end{equation}}
\newcommand{\beq}{\begin{eqnarray}}
\newcommand{\eeq}{\end{eqnarray}}
\def\ba{\begin{align}}
\def\ea{\end{align}}

\newcommand{\dd}{\mathrm{d}}

\newcommand{\fd}[1]{\textcolor{cyan}{\sf{[F: #1]}}}
\newcommand{\cm}[1]{\textcolor{orange}{\sf{[Caio: #1]}}}

\begin{document}
	
\title{Extreme-mass-ratio inspirals in ultra-light dark matter}

\author{Francisco Duque}
%
\affiliation{Max Planck Institute for Gravitational Physics (Albert Einstein Institute) Am Mühlenberg 1, D-14476 Potsdam, Germany}

\affiliation{CENTRA, Departamento de F\'{\i}sica, Instituto Superior T\'ecnico -- IST, Universidade de Lisboa -- UL,
	Avenida Rovisco Pais 1, 1049 Lisboa, Portugal}

\author{Caio F. B. Macedo}
\affiliation{Faculdade de Física, Campus Salin\'opolis, Universidade Federal do Par\'a, 68721-000, Salin\'opolis, Par\'a, Brazil}

\author{Rodrigo Vicente}
\affiliation{Institut de Fisica d’Altes Energies (IFAE), The Barcelona Institute of Science and Technology, Campus UAB, 08193 Bellaterra (Barcelona), Spain}

\author{Vitor Cardoso}
\affiliation{CENTRA, Departamento de F\'{\i}sica, Instituto Superior T\'ecnico -- IST, Universidade de Lisboa -- UL,
	Avenida Rovisco Pais 1, 1049 Lisboa, Portugal}
\affiliation{Niels Bohr International Academy, Niels Bohr Institute, Blegdamsvej 17, 2100 Copenhagen, Denmark}
\affiliation{Yukawa Institute for Theoretical Physics, Kyoto University, Kyoto
606-8502, Japan}
\date{\today}%

\begin{abstract}
Previous works have argued that future gravitational-wave detectors will be able to probe the properties of astrophysical environments where binaries coalesce, including accretion disks, but also dark matter structures. Most analyses have resorted to a Newtonian modelling of the environmental effects, which are \emph{not suited} to study extreme-mass-ratio inspirals immersed in structures of ultra-light bosons.   
In this letter, we use relativistic perturbation theory to consistently study these systems in spherical symmetry. We compute the flux of scalar particles and the rate at which orbital energy is dissipated via gravitational radiation and depletion of scalars, i.e. dynamical friction. Our results confirm that the Laser Interferometer Space Antenna will be able to probe ultra-light dark matter structures by tracking the phase of extreme-mass-ratio inspirals.
\end{abstract}

\maketitle

\noindent{\bf{\em Introduction.}}
Progress of gravitational-wave (GW) astronomy requires modeling increasingly complex sources. Extreme-mass-ratio inspirals (EMRIs) -- to be observed by the space-borne Laser Interferometer Space Antenna (LISA) mission~\cite{2017arXiv170200786A} -- correspond to binaries where a stellar-mass compact object orbits around a supermassive black hole (SMBH) with $\gtrsim 10^5 \, M_\odot$. EMRIs perform $\sim 10^5$ orbital cycles while in the LISA frequency band~\cite{Hinderer:2008dm, Babak:2017tow, Barack:2018yly}. 
They hold a vast scientific potential~\cite{LISA:2022yao, LISACosmologyWorkingGroup:2022jok, LISA:2022kgy}, but require the construction of extremely accurate waveforms~\cite{Hughes:2021exa, Wardell:2021fyy}.

EMRIs may evolve in non-vacuum environments, such as active galactic nuclei disks~\cite{Barausse:2014tra,Pan:2021ksp, Pan:2021oob, Derdzinski:2022ltb, LIGOScientific:2020stg, Abbott:2020khf, Abbott:2020tfl, Abbott:2020uma, Tagawa:2019osr,Tagawa:2020qll, Derdzinski:2022ltb}, and should be accreting matter throughout the evolution. The motion of the smaller body generates trailing density wakes in the surrounding medium, acting as dynamical friction that accelerates the inspiral~\cite{1943ApJ....97..255C, Ostriker:1998fa}. Environmental effects occur in accretion disks~\cite{Shakura:1972te, Abramowicz:2011xu} or in galactic dark matter (DM) structures~\cite{1980ApJ...238..471R, Navarro:1995iw, 1990ApJ...356..359H, PhysRevD.89.104059, Cardoso:2019rou, Zwick:2022dih, CanevaSantoro:2023aol}.
Studies on their relevance for GW astronomy often resort to Newtonian approximations~\cite{Kocsis:2011dr, Kavanagh:2020cfn, Coogan:2021uqv,Cole:2022yzw, Garg:2022nko, Tomaselli:2023ysb, Berezhiani:2023vlo}, but EMRIs inherently probe strong-field gravity~\cite{Vicente:2022ivh, Traykova:2021dua,  Traykova:2023qyv, Speeney:2022ryg, Speri:2022upm, Derdzinski:2020wlw,2002ApJ...565.1257T, Derdzinski:2020wlw}.
Recently, a fully-relativistic framework was developed to study GW emission by EMRIs in spherically symmetric, but otherwise generic, spacetimes~\cite{Cardoso:2021wlq, Cardoso:2022whc, 1967ApJ...149..591T, 1985ApJ...292...12D, PhysRevD.46.4289, Allen:1997xj, Sarbach:2001qq, PhysRevD.71.104003, PhysRevD.69.044025}. Relevant differences were found in the GW flux with respect to vacuum for EMRIs immersed in a fluid-like DM halo~\cite{Cardoso:2021wlq,Cardoso:2022whc}. 

In this Letter, we extend the analysis to axionic environments of ultra-light (scalar) fields. New degrees of freedom with mass $m_\Phi \sim 10^{-6}$--$10^{-23} \, \text{eV}$ appear naturally in extensions of the Standard Model~\cite{PhysRevLett.38.1440, PhysRevLett.40.223, Arvanitaki:2009fg} and have been suggested as a component of DM~\cite{Hui:2016ltb, Hui:2021tkt, 1985MNRAS.215..575K}. They can form self-gravitating structures~\cite{PhysRev.172.1331, Seidel:1993zk, Lee:1995af, Liebling:2012fv} (known as boson stars or DM solitons) which describe well the core of DM halos when the bosons are ``fuzzy'' (i.e., their mass is~$m_\Phi \sim 10^{-22}$--$10^{-19} \, \text{eV}$)~\cite{Torres:2000dw, Schive:2014dra, Schive:2014hza,Veltmaat:2018dfz, Barranco:2017aes, Cardoso:2022nzc}. Here, their de Broglie wavelength is $\lesssim \text{kpc}$, and they behave as waves at these (and smaller) length scales, alleviating tensions with some galactic observations~\cite{1994Natur.370..629M, Navarro:1996gj, Weinberg:2013aya}. When their Compton wavelength is comparable to the size of a spinning BH, they can efficiently extract rotational energy from it via superradiance~\cite{Brito:2015oca, 1972Natur.238..211P, Dolan:2007mj, East:2017ovw,East:2018glu}, and condensate into boson clouds~\cite{Brito:2015oca,Detweiler:1980uk,Cardoso:2005vk,Arvanitaki:2010sy,Brito:2014wla,Baumann:2019eav}. Other mechanisms to form boson condensates include accretion~\cite{Hui:2019aqm,Clough:2019jpm,Cardoso:2022nzc} or dynamical capture from a DM halo~\cite{Budker:2023sex}.


\noindent{\bf{\em Theory.}}
%
Consider a minimally coupled complex scalar~$\Phi$ described by the action (we use the metric signature $(-+++)$ and geometrized units $G=c=1$)
\begin{nofleqn}{equation}
    S=\int \dd^4 x \sqrt{-g}\big(\tfrac{R}{16 \pi}-\tfrac{1}{2}\partial_\mu \Phi\partial^\mu \Phi^*-\tfrac{1}{2}V(\Phi^* \Phi)+ \mathcal{L}^\mathrm{m}\big) , 
\end{nofleqn}
with~$R$ the Ricci scalar,~$V$ a self-interaction potential, and $\mathcal{L}^\mathrm{m}$ describes the matter sector, which interacts with the scalar only via the minimal coupling to the metric~$g_{\mu \nu}$.
This theory has a global~$\mathrm{U}(1)$ symmetry with associated Noether current and charge, respectively,
\beq
    J_Q^\mu=2 g^{\mu\nu} \text{Im} \left[\Phi^* \partial_\nu \Phi\right], \quad Q= \int \dd^3x \sqrt{-g}J_Q^t.
\eeq

The action above results in the Einstein-Klein-Gordon (EKG) system
\beq
	G_{\mu \nu}=8\pi T_{\mu \nu}, \qquad \Box_g\Phi=\frac{\partial V}{\partial \Phi^*}, 
	 \label{eq:EKG}
\eeq
where~$\Box_g\equiv \partial_\mu\left[\sqrt{-g}\,\partial^\mu\left(\cdot\right)\right]/\sqrt{-g}$, plus a set of equations for the matter sector. Here,~$G_{\mu \nu}$ is the Einstein tensor and~$T_{\mu \nu}$ is the total energy-momentum tensor
\begin{equation}
T_{\mu \nu}\equiv T^\Phi_{\mu \nu}+T^\textrm{m}_{\mu \nu},
\end{equation}
with $T^\Phi_{\mu \nu}= \nabla_{(\mu} \Phi^* \nabla_{\nu)} \Phi-\frac{g_{\mu \nu}}{2}(\nabla^\alpha \Phi^*\nabla_\alpha \Phi+V)$ the scalar field contribution.
We focus on sufficiently dilute environments such that~$V\approx \mu^2 |\Phi|^2$, with~$\mu = m_\Phi/\hbar$ the inverse Compton wavelength of the scalar.

\noindent{\bf{\em EMRI system.}}
%
%
The matter sector 
is the smaller object in the EMRI, modeled by a point particle of mass~$m_p$ with Lagrangian density
\beq
    \mathcal{L}^\textrm{m}= m_{p} \int \dd \tau \tfrac{1}{\sqrt{-g}}\,\delta^{(4)}\left[x-x_p(\tau)\right],
\eeq
where $x_p(\tau)$ is the particle worldline parameterized by its proper time~$\tau$. 
The corresponding energy-momentum tensor is then~$T_\textrm{m}^{\mu \nu}= m_{p} \int  \dd\tau \frac{1}{\sqrt{-g}}\delta^{(4)}\left[x-x_p(\tau)\right] \dot{x}_p^\mu \dot{x}_p^\nu$, where~$\dot{x}_p^\mu\equiv \dd x_p^\mu/\dd\tau$.
Treating the secondary as a point particle is a good approximation since the scalar field is light enough to be insensitive to the internal structure of these objects. More precisely, we consider scalars with de Broglie wavelength much larger than the size of the secondary object~\cite{Vicente:2022ivh,Traykova:2021dua,Traykova:2023qyv}.

\noindent{\bf{\em Background solution.}}
%
The particle acts as a source of perturbations to a given background solution~$\{\widehat{g}_{\mu \nu},\widehat{\Phi}\}$ of the EKG system~\eqref{eq:EKG}, with $T^\text{m}_{\mu\nu} = 0$. 
The background spacetime is taken to be static and spherically symmetric
\begin{nofleqn}{equation}
    \dd\widehat{s}^2 \equiv \widehat{g}_{\mu \nu}\dd x^\mu \dd x^\nu\approx -A(r) \mathrm{d}t^2+\frac{\mathrm{d}r^2}{B(r)}+r^2  \mathrm{d}\Omega^2,
    \label{eq:MetricGenProfile}
\end{nofleqn}
with~$\dd\Omega^2 = \dd \theta^2 + \sin^2 \theta \dd\varphi^2$ the line element on the 2-sphere. The associated background field has the form
\begin{equation}
	\widehat{\Phi} \approx \phi_0(r) e^{-i \omega t} \,, \quad \omega \in \mathbb{R}. \label{eq:ScalarGenProfile}
\end{equation}
This background is used to describe a BH of mass~$M_\mathrm{BH}$ dressed with a scalar configuration with mass~$M_\Phi \equiv\int\dd^3 x\sqrt{-g}\, T^{\Phi}_{t t}/A$, which is bound either by the BH gravity ($M_\mathrm{BH}\gg M_\Phi$) or by its own self-gravity ($M_\mathrm{BH}\ll M_\Phi$)~\cite{Barranco:2017aes}. These two limits correspond, respectively, to boson clouds and DM solitons doped with a parasitic BH.
From no-hair results~\cite{Ruffini:1971bza,Bekenstein:1972ny}, static and spherically symmetric solutions cannot exist, and the scalar field must be swallowed by the BH. However, for ultra-light scalars the accretion timescale can be much larger than the EMRI inspiral (or than an Hubble time~\cite{Cardoso:2022nzc}), and Eqs.~\eqref{eq:MetricGenProfile} and~\eqref{eq:ScalarGenProfile} are a good approximation to the background solution. 

\noindent{\bf{\em Linear perturbations.}}
%
The point particle sources small perturbations to the background solution, resulting in the spacetime metric~$g_{\mu \nu}$ and scalar field~$\Phi$:
\beq
g_{\mu\nu} \approx \widehat{g}_{\mu\nu}+ q\, \delta g_{\mu\nu}, \qquad \Phi \approx \widehat{\Phi} + q\, \delta\Phi \, ,
\eeq
accurate up to~$\mathcal{O}(q^2)$, with~$q\equiv m_p/M_\mathrm{BH}$. Spherical symmetry allows us to decompose any perturbation into irreducible representations of~$\text{SO}(3)$, grouped into polar and axial sectors, depending on their behaviour under parity transformations~\cite{Regge:1957td, Zerilli:1970wzz, Sago:2002fe, Detweiler:2003ci, Martel:2005ir, NIST:DLMF}. We focus on the polar sector, which dominates GW emission and couples scalar to gravitational perturbations. 
In the \emph{Supplemental Material} (SM), we describe how to reduce the system to a set of ``wave-like'' partial differential equations, which we solve numerically using techniques previously applied in EMRI modelling~\cite{Cardoso:2022whc, Macedo:2013jja}. 

At leading order in $q$, the particle follows geodesics of the background spacetime~$\widehat{g}_{\mu \nu}$. For simplicity, we focus on \emph{quasi-circular} EMRIs, with particle worldline $x_p^\mu(t)=\big(\varepsilon_{\rm p}/A_p,r_p,\pi/2,\Omega_p t\big)$, with $A_p\equiv A(r_p)$, and where $\varepsilon_{\rm p}\equiv A_p/\sqrt{\smash[b]{A_p-r_p^2\Omega_p^2}}$ is the particle energy per unit mass and $\Omega_p\equiv \sqrt{A'(r_p)/2r_p}$ the orbital angular frequency~\cite{Cardoso:2008bp}. 

\noindent{\bf{\em Adiabatic inspiral and fluxes.}}
%
As the secondary orbits the central BH, it transfers part of its energy to the scalar environment and to GWs. For small $q$, the energy dissipated over an orbit is much smaller than the orbital energy, and the secondary evolves \emph{adiabatically} over a succession of geodesics~\cite{Hughes:2021exa, Barack:2018yvs}.
To follow its evolution, we only need to compute fluxes in the gravitational and scalar sectors using the linearized EKG system. 

We compute energy fluxes at the leading order~$O(q^2)$ using~$T^\Phi_{\mu \nu}$ and Isaacson's effective energy-momentum tensor~$T^g_{\mu \nu}$~\cite{Isaacson:1968zza},
\beq
    \dot{E}^{S}_{L} =-\sigma_L \lim_{r\rightarrow r_L} r^2 \sqrt{A B}\int \dd\Omega \,  T_{tr}^{S}, 
\eeq
with sector~$S=\{\Phi,g\}$, location~$L=\{H,\infty\}$ (at horizon and spatial infinity, respectively), and~$\sigma_{H,\infty}=-,+$. Note that the energy of a non-relativistic scalar field environment is~$M_\Phi \approx  Q\, \omega$, and that the dissipation process results in~$\dot{M}_\Phi \approx \dot{Q}\, \omega $. Thus, the fluxes of Noether charge
\beq
    \dot{Q}_{L} =\sigma_L \lim_{r\rightarrow r_L} r^2 \sqrt{\frac{A}{B}}\int \dd\Omega \,  J_Q^r,
\eeq
are necessary to compute the rate of change of the particle energy, $\dot\varepsilon_{\rm p}\equiv \sum_{S,L,\ell} {}^\ell\dot\varepsilon_{\rm p,L}^{\,S}$, with~$\ell$ the azimuthal number and~\cite{Annulli:2020lyc} 
\beq
{}^\ell\dot\varepsilon_{\rm p,L}^{\,\Phi}  \approx -\frac{{}^\ell\dot{E}^{\Phi}_{L}- {}^\ell\dot{Q}_L\omega}{m_p},\quad {}^\ell\dot\varepsilon_{\rm p,L}^{\,g} = -\frac{{}^\ell\dot{E}^{g}_{L}}{m_p}\,. 
\eeq
The flux expressions in terms of the perturbation asymptotic amplitudes are given in the SM.

\noindent{\bf{\em Fuzzy DM soliton.}}
%
Let us consider first isolated self-gravitating solutions of the EKG system~\cite{PhysRev.172.1331,Ruffini:1971bza,Liebling:2012fv}. We focus on \emph{Newtonian} boson stars (NBSs), where~$|\omega-\mu| \ll \mu$~\cite{Annulli:2020lyc, 2011PhRvD..84d3531C, Kling:2017mif, Chan:2023crj}, which may describe the inner core of DM halos~\cite{Hui:2016ltb,Hui:2021tkt}. They follow the relation~\cite{2011PhRvD..84d3531C, Boskovic:2018rub,Annulli:2020lyc}
\beq
M_\Phi	\approx 9 \times  10^{8}M_\odot \Big(\frac{0.1\, \text{kpc}}{R_\text{98}}\Big) m_{22}^{-2} ,
\eeq
with~$m_{22}\equiv m_\Phi/10^{-22}\,\mathrm{eV}$ and $R_{98}$ the radius enclosing $98\%$ of~$M_\Phi$ (reference values for a Milky-Way-like galaxy~\cite{DeMartino:2018zkx}). Their oscillating frequency is~$\omega/\mu \approx 1- 0.163\left(M_\Phi \mu\right)^2$~\cite{Annulli:2020lyc,Guzman:2004wj},
where~$\mu \approx4.79\,\text{yr}^{-1} \, m_{22}$, and~$M_\Phi \mu \sim 10^{-3}\,m_{22} (M_\Phi/10^9 M_\odot)$. The metric functions are $A \approx B\approx 1 + 2 U_\mathrm{NBS}(r)$, where~$U_\mathrm{NBS}$ is the gravitational potential of the NBS, satisfying~$\nabla^2 U_\mathrm{NBS}=\mu^2\phi_{0, \mathrm{NBS}}^2$. Accurate fits for~$U_{\rm NBS}$ and the radial distribution~$\phi_{0, \rm{NBS}}$ are known~\cite{Annulli:2020lyc} (for details see SM).

The dynamics of NBSs in the \emph{non-relativistic} regime was studied in Ref.~\cite{Annulli:2020lyc}, including dissipation via scalar field depletion by quasi-circular binaries with~$\Omega_p\ll \mu$. There, it was shown that the scalar fluxes are well described by analytic expressions [cf. Eqs.~(147)-(149), or our SM]. Here, for configurations with~$M_{\Phi} \mu \lesssim 0.25$, our numerical results for the scalar field fluxes -- obtained from solving the linearized \textit{fully relativistic} EKG system -- agree with these expressions, with relative errors~$\lesssim 10 \% $,  within their range of validity. 
Even for more compact configurations, they capture the correct dependence of the fluxes on the orbital frequency.

For~$\ell$-mode scalar field perturbations to be depleted the orbital frequency must satisfy $\ell \Omega_p > \mu-\omega$, otherwise they are confined. If the orbital dynamics is solely determined by the NBS gravity, it can be shown that~$\Omega_p \lesssim \mu-\omega$: the orbital frequency is always too small to generate dipolar ($\ell = 1$) radiation. However, dipolar depletion is possible for more compact relativistic configurations, or in a more realistic description of fuzzy DM cores that includes a central SMBH. 
Any EMRI observable by LISA (and most near-future GW detectors) must have an orbital frequency~$\Omega_p \gg \mu \sim 10^{-7}m_{22}$\,{\rm Hz}, thus sourcing \emph{relativistic} perturbations. 
Such orbital frequencies \emph{cannot} be supported by the NBS gravity alone, but they are allowed in compact boson stars~\cite{Macedo:2013jja}, or in NBSs hosting a parasitic massive BH, to which we now turn.

\noindent{\bf{\em Fuzzy DM soliton hosting a parasitic BH.}}
Large galaxies (like the Milky Way) are expected to host a SMBH~\cite{Kormendy:1995er,Kormendy:2013dxa}. To describe the actual galactic center, we consider a SMBH at the center of a NBS. As mentioned, any spherical scalar configuration around a BH is a transient. However, for an ultra-light boson mass~$m_{22}\lesssim 800$, a SMBH with~$M_{\rm BH}\lesssim 10^6 M_\odot$ at the center of a soliton core of a DM halo with mass~$M_{\rm halo} \lesssim 10^{15} M_\odot$ takes more than an Hubble time to accrete $10\%$ of the soliton mass~\cite{Cardoso:2022nzc}. 
We can then safely neglect accretion during the observation timescale of an EMRI evolving in these environments.

\begin{figure}[t]
\centering \includegraphics[width=\linewidth]{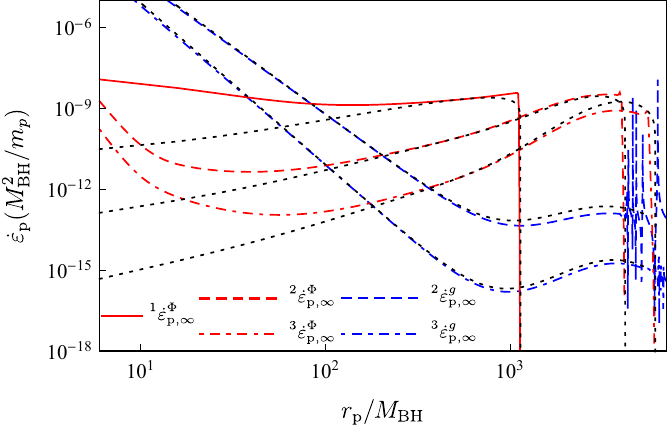}
\caption{Energy loss rate in the gravitational ($|{}^\ell\dot\varepsilon_{\rm p,\infty}^{\,g}|$) and scalar sectors ($|{}^\ell \dot{\epsilon}_{p,\infty}^{\,\Phi}|$) for an EMRI in a NBS with $M_{\Phi}\mu = 0.23$ hosting a parasitic BH of mass $M_{\rm BH} = 0.02M_{\Phi}$; the soliton size is~$R_{98}\sim 8 \times 10^3 M_{\rm BH} $. Scalar modes are well described by low-frequency analytic expressions from~\cite{Annulli:2020lyc}
in their regime of validity, and the gravitational contributions follow the vacuum analytic expressions for quadrupole and octupole
(all shown as black dots).  
}   
    \label{fig:BosonStar}
\end{figure}
As shown in SM, if~$M_{\Phi} \gg M_{\rm BH}$ and~$M_{\Phi} \mu \ll 1  $, the background spacetime can be described by
\begin{nofleqn}{equation}
A \approx  \Big(1- \frac{2 M_\text{BH}}{r}\Big) e^{2 U_{\rm NBS} (r)} \, , \quad B \approx 1-\frac{2m(r)}{r} \,,
\end{nofleqn}
with
\begin{gather}
\phi_0 \approx \phi_{0, \rm NBS}(r) e^{-2i \mu M_{\rm BH} \log\left(1 - \frac{2M_{\rm BH}}{r} \right)}\, , \\
m  \equiv  M_\text{BH} + 4\pi \mu^2 \int_{2M}^r \dd r' r'^2 [\phi_{0, \rm{NBS}}(r')]^2\,,
\end{gather}
with~$U_{\rm NBS}$ and~$\phi_{0, \rm NBS}$ the (isolated) NBS profiles. Note that the central BH dominates the dynamics close to the horizon, where $m(r) \approx M_{\rm BH}$ and $U_{\rm NBS}(r) \approx - \frac{1}{2}   M_{\Phi} / R_{98}\ll1$~\cite{Cardoso:2022nzc}. There, the background spacetime reduces to Schwarzschild with a redshifted time coordinate controlled by the NBS compactness, $A \approx \left(1 - 2M_{\rm BH} /r \right)\left( 1- M_{\Phi} / R_{98} \right)$. This behaviour was noted for other extended matter distributions around BHs~\cite{Cardoso:2021wlq, Figueiredo:2023gas,Brito:2023pyl} and, as we show in SM, it happens generically when the typical length scale of a Newtonian matter distribution is much larger than the BH size.
Far from the horizon, the NBS gravity dominates the dynamics and, from~$U_{\rm NBS}\ll 1$, we recover the linear superposition of the two Newtonian potentials~$A \approx 1 + 2\left(U_{\rm NBS}-M_{\rm BH}/ r\right)$.
The innermost stable circular orbit (ISCO) frequency is $\Omega_{\rm ISCO}/\mu \approx (M_{\rm BH} \mu)^{-1}(1-M_\Phi/2R_{98})/6\sqrt{6}$ ($\gg 1$ for ultra-light fields), implying that EMRIs in this background can generate relativistic perturbations and dipolar depletion while in the LISA frequency band.

The contributions of scalar field depletion and gravitational radiation to the energy loss rate of an EMRI are shown in Fig.~\ref{fig:BosonStar}, as a function of the distance to the central BH. For illustration purposes, we considered a soliton of mass~$M_{\Phi} \mu = 0.23$
hosting a parasitic SMBH with~$M_{\rm BH}=0.02 M_\Phi$; this would accrete the soliton on a timescale~$\sim 10^5\, \mathrm{yr} \,m_{22}^{-1}$~\cite{Cardoso:2022nzc}, much larger than the inspiralling timescale or the period of observation for an EMRI in LISA and a boson with~$m_{22}\lesssim 10^2$. 


For low-frequency EMRIs, the fluxes are well described by the analytical predictions both in the scalar~\cite{Annulli:2020lyc} and gravitational sectors, apart from the resonant behavior in the gravitational sector for $\Omega_p \lesssim (\mu-\omega)/\ell$, which is a relativistic feature~\cite{Macedo:2013jja}.  
The energy dissipated through scalar depletion for orbital frequencies~$\Omega_p\gg \mu \sim 10^{-7}m_{22}\,{\rm Hz}$ can be much larger 
than the low-frequency predictions, since relativistic corrections become active.
The contribution from dipolar scalar field depletion 
-- which dominates the energy loss in the scalar channel for all orbital radii where it is active -- 
shows a slightly different behaviour than higher-$\ell$ modes, possibly due to its exceptional character (there is no gravitational radiation for $\ell=1$).

Fluxes at the BH horizon are several orders of magnitude smaller than at infinity. For low orbital frequencies~$\Omega_p\lesssim 16.3/\ell\, (M_\Phi \mu)^2 m_{22} {\rm \, nHz}$ the normal modes of the bosonic configuration are resonantly excited by the EMRI~\cite{Macedo:2013jja,Annulli:2020lyc,Chan:2023crj}. 
Even though these scalar perturbations are bound to the soliton, the corresponding GWs carry information about the global structure of the background. The stirring of the bosonic profile by the EMRI at such frequencies also induces resonant accretion, but its contribution to the energy loss rate is orders of magnitude smaller than the flux of energy in GWs at infinity (see Fig.~2 of SM).

\noindent{\bf{\em Boson cloud.}}
%
%
Finally, we apply our framework to an EMRI evolving in a \emph{spherical} (ground-state) boson cloud, a solution of the EKG system where the scalar field structure is sustained by the BH gravity alone. For diluted clouds one may neglect its backreaction on the geometry. The background spacetime is then taken to be (approximately) described by the Schwarzschild metric~$A \approx B \approx 1-2M_{\rm BH}/r$, whereas the scalar field profile and oscillating frequency are~\cite{Brito:2015oca,Detweiler:1980uk,Cardoso:2005vk,Arvanitaki:2010sy,Baumann:2019eav,Baumann:2019ztm}
\begin{equation}
\phi_0 \approx C e^{-M_{\rm BH} \mu^2 r}  e^{-2i \mu M_{\rm BH} \log\left(1 - \frac{2M_{\rm BH}}{r} \right) }, \label{eq:BCNewtonianApprox}
\end{equation}
and~$\omega/\mu \approx 1- \tfrac{1}{2}(M_{\rm BH} \mu)^2$, with~$C\equiv \sqrt{\frac{M_\Phi}{\pi M_{\rm BH}}}(M_{\rm BH} \mu)^2$.
Superradiance provides an effective mechanism for formation of these structures~\cite{Brito:2014wla, Brito:2015oca}, which can lead to the growth of clouds with~$M_\Phi\approx 0.1 M_{\rm BH}$ on short timescales~\cite{East:2017ovw,East:2018glu}. This formation mechanism requires the seed BH to be spinning, and superradiant clouds are born with angular momentum. Thus, they cannot possibly be spherically symmetric. Even though BH spin is key for the growth of these structures, it plays a less crucial role in \emph{some} of its physics. Spherical boson clouds can be grown via other mechanisms, like accretion~\cite{Hui:2019aqm,Clough:2019jpm,Cardoso:2022nzc} or dynamical capture from a DM halo~\cite{Budker:2023sex}.

\begin{figure}[t]
    \centering \includegraphics[width=\columnwidth]{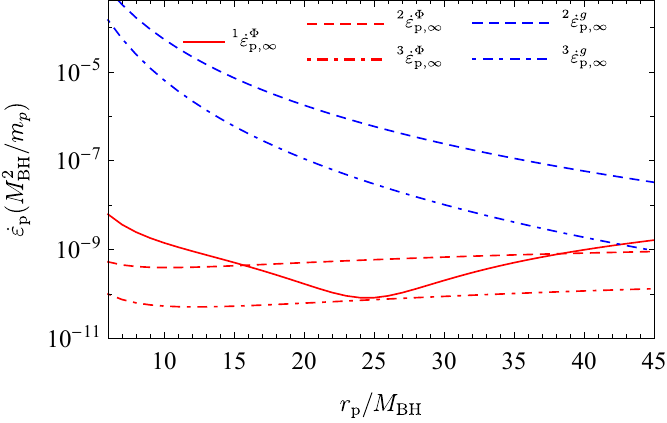}
    \caption{Energy loss rate due to the fluxes at infinity in the gravitational ($|{}^\ell\dot\varepsilon_{\rm p,\infty}^{\,g}|$) and scalar sectors ($|{}^\ell \dot{\epsilon}_{p,\infty}^{\,\Phi}|$) for an EMRI evolving in a boson cloud with~$M_{\rm BH}\mu = 0.08$ and~$M_{\Phi}=0.1M_{\rm BH}$; the cloud size is~$R_{98} \sim 6 \times 10^2 M_{\rm BH}$. For diluted clouds the test field approximation is valid and the energy dissipated via scalar depletion is linear in~$M_\Phi$.} 
    \label{fig:Cloud}
\end{figure}

Figure~\ref{fig:Cloud} shows the dependence of scalar and GW emission on the orbital radius for an EMRI with~$M_{\rm BH} \mu =0.08$ and $M_{\rm \Phi}=0.1M_{\rm BH}$.  
Our results are in qualitative agreement with those of Ref.~\cite{Brito:2023pyl}. The small discrepancies are due to differences in the scalar field profile,~$\phi_{0}$, from the Newtonian approximation employed (c.f., \eqref{eq:BCNewtonianApprox}), and the relativistic (numerical) profile used in Ref.~\cite{Brito:2023pyl}. The effect from scalar depletion dominates over the one of gravitational radiation-reaction for sufficiently large orbital radii. Our results suggest that dissipation via dipolar scalar depletion is especially sensitive to the scalar field profile of the background (compare Figs.~\ref{fig:BosonStar} and~\ref{fig:Cloud}).  

%
%

\begin{figure}
    \centering
    \includegraphics[width=\linewidth]{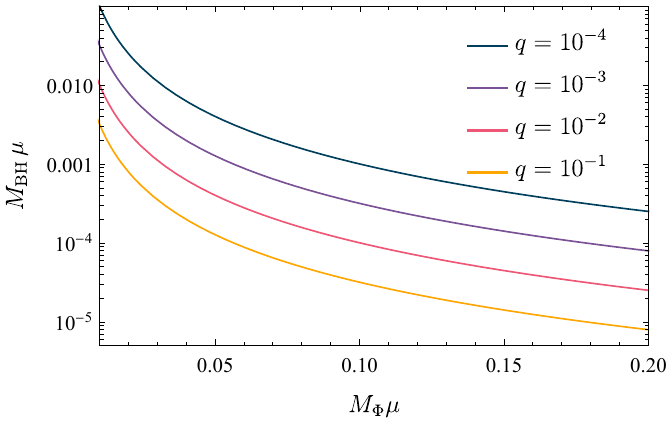}
    \caption{Dephasing of GWs from an EMRI in a DM soliton. Each line stands for a one-cycle dephasing. Above the line the dephasing is larger than one cycle.}
    \label{fig:dephasing}
\end{figure}

\noindent{\bf{\em Observational prospects.}}
%
The dephasing of an EMRI gravitational waveform can be roughly estimated by $\delta \psi/(2\pi) \sim \Omega_p T_{\rm obs}^2/(\pi T_\Phi)$, where~$T_\Phi\sim \epsilon_p/\dot{\epsilon}_p$~\cite{Kocsis:2011dr}. In Fig.~\ref{fig:dephasing}, choosing~$T_{\rm obs}=4\, \mathrm{yr}$, we show the region in the parameter space where the dephasing of signal due to the fuzzy DM soliton environment is larger than one cycle. This value is typically used as a rule of thumb for measurability of beyond-vacuum GR effects with LISA~\cite{Kocsis:2011dr}. 
We estimate~$\dot{\varepsilon}_p$ by fitting the numerical data in the LISA frequency band, in the range $M_\Phi\mu=[0.1,0.25]$ and $M_\text{BH}/M_\Phi=[0.01,0.04]$. For $r_p<15M_{\rm BH}$, the fit underestimates the flux, so our results for the dephasing are conservative. 
For a fixed mass-ratio~$q$, the dephasing due to the scalar environment surpasses the criteria threshold provided that~$M_\Phi \mu$ and~$M_{\rm BH} \mu$ are \emph{large} enough. While some cosmological simulations~\cite{Schive:2014dra,Schive:2014hza,Veltmaat:2018dfz} indicate that~$M_\Phi \mu \sim 10^{-3}$ for low-redshift halos of mass~$M_h\sim 10^{12} M_\odot$, there is considerable dispersion in the literature (most probably due to tidal stripping events in the halos history~\cite{Chan:2021bja}) allowing for~$M_\Phi \mu \lesssim 0.02\, \sqrt{m_{22}}$. This makes the effect potentially observable for sufficiently compact field configurations of scalars with~$m_\Phi\gtrsim 10^{-20}{\rm eV}$. For boson clouds, we use directly the numerical results and find that for clouds with mass $M_\Phi/M_\text{BH} \gtrsim 0.01$, any scalar field mass $M_\Phi \mu \gtrsim 0.08$ leads to a 1 cycle dephasing for a  mass-ratio $ q \gtrsim 5\times 10^{-5}$. 


\noindent{\bf{\em Discussion.}}
%
Our results highlight the importance of relativistic modelling of environmental effects. Although we recover several features previously discussed in a Newtonian context~\cite{Barausse:2014tra,Cardoso:2019rou,Baumann:2019eav, Baumann:2019ztm,Cole:2022yzw,Tomaselli:2023ysb}, others can only be captured within a fully relativistic framework, such as the behavior of the dipolar mode or the resonant accretion by the central BH.

Comparison of Figs.~\ref{fig:BosonStar} and~\ref{fig:Cloud} suggests future GW observations can probe and discriminate between different ultra-light dark matter distributions, as the energy flux in the scalar sector is sensitive to the profile of the background scalar distribution. Although we focused the discussion on EMRIs for LISA, our results can be relevant for SMBH binaries in the Pulsar Timing Array (PTA) frequency-band ($\sim 10^{-9}-10^{-7}\, \text{Hz}$), as the dissipation via scalar depletion dominates over GW radiation reaction for large orbital separations~\cite{ NANOGrav:2023hfp, NANOGrav:2023hvm, NANOGrav:2023gor, EPTA:2023fyk, EuropeanPulsarTimingArray:2023qbc, Antoniadis:2023zhi, Ghoshal:2023fhh, Ellis:2023dgf}. 
Such prospects motivate further explorations, including a systematic analysis of different ultra-light DM configurations. This could assist in finding semi-analytic descriptions for the energy loss via scalar depletion (i.e. dynamical friction) that would facilitate the implementation of these environmental effects in state-of-the-art EMRI waveform models, like the modular FEW package~\cite{Hughes:2021exa, Katz:2021yft, Speri:2023jte}.
Another important extension is the study of more generic orbits, as dynamical friction does not necessarily circularize the orbits. Non-vacuum environments are known to lead to eccentricity excitation~\cite{Nelson:2005xh, Cardoso:2020iji,Tomaselli:2023ysb, Rahman:2023sof}, and DM distributions at galactic centres can induce orbital precession that competes with vacuum GR~\cite{Destounis:2022obl, Destounis:2023khj}. Finally, our results can impact binary merger rates, similarly to EMRI formation assisted by gas flows in accretion disks, a mechanism which has been suggested to dominate the fraction of detectable EMRIs by LISA~\cite{Pan:2021ksp}.

\noindent{\bf{\em Acknowledgements.}}
We are thankful to Richard Brito for very fruitful discussions. 
F.D. acknowledges financial support provided by FCT/Portugal through grant No. SFRH/BD/143657/2019.
C.F.B.M. would like
to thank Fundação Amazônia de Amparo a Estudos
e Pesquisas (FAPESPA), Conselho Nacional de Desenvolvimento Científico e Tecnológico (CNPq) and Coordenação de Aperfeiçoamento de Pessoal de Nível Superior (CAPES) – Finance Code 001, from Brazil, for partial financial support. 
R.V. is supported by grant no. FJC2021-046551-I funded by MCIN/AEI/10.13039/501100011033 and by the European Union NextGenerationEU/PRTR. R.V. also acknowledges the support from the Departament de Recerca i Universitats from Generalitat de Catalunya to the Grup de Recerca `Grup de F\'isica Te\`orica UAB/IFAE' (Codi: 2021 SGR 00649).
We acknowledge support by VILLUM Foundation (grant no. VIL37766) and the DNRF Chair program (grant no. DNRF162) by the Danish National Research Foundation.
V.C.\ is a Villum Investigator and a DNRF Chair.  
V.C. acknowledges financial support provided under the European Union’s H2020 ERC Advanced Grant “Black holes: gravitational engines of discovery” grant agreement no. Gravitas–101052587. 
Views and opinions expressed are however those of the author only and do not necessarily reflect those of the European Union or the European Research Council. Neither the European Union nor the granting authority can be held responsible for them.
This project has received funding from the European Union's Horizon 2020 research and innovation programme under the Marie Sklodowska-Curie grant agreement No 101007855 and No 101131233.

\bibliography{References}

\end{document}


\title{Supplemental Material to ``Axion weak leaks: extreme mass-ratio inspirals in ultralight dark matter''}

%
\author{Francisco Duque}
\affiliation{Max Planck Institute for Gravitational Physics (Albert Einstein Institute) Am Mühlenberg 1, D-14476 Potsdam, Germany}

\affiliation{CENTRA, Departamento de F\'{\i}sica, Instituto Superior T\'ecnico -- IST, Universidade de Lisboa -- UL,
	Avenida Rovisco Pais 1, 1049 Lisboa, Portugal}
\author{Caio F. B. Macedo}
\affiliation{Faculdade de Física, Campus Salin\'opolis, Universidade Federal do Par\'a, 68721-000, Salin\'opolis, Par\'a, Brazil}
\author{Rodrigo Vicente}
\affiliation{Institut de Fisica d’Altes Energies (IFAE), The Barcelona Institute of Science and Technology, Campus UAB, 08193 Bellaterra (Barcelona), Spain}
\author{Vitor Cardoso}
\affiliation{CENTRA, Departamento de F\'{\i}sica, Instituto Superior T\'ecnico -- IST, Universidade de Lisboa -- UL,
	Avenida Rovisco Pais 1, 1049 Lisboa, Portugal}
\affiliation{Niels Bohr International Academy, Niels Bohr Institute, Blegdamsvej 17, 2100 Copenhagen, Denmark}

\maketitle 

In this Supplemental Material, we give a detailed derivation of the evolution equations for the gravitational and scalar field linear perturbations in spherically-symmetric, but otherwise generic, backgrounds. Our procedure is based on previous works on relativistic stars~\cite{Kojima:1992ie, Allen:1997xj}, and our own studies of extreme mass-ratio inspirals (EMRIs) immersed in dark-matter (DM) halos constituted by an anisotropic fluid~\cite{Cardoso:2021wlq, Cardoso:2022whc}. We also explicitly express the different fluxes in terms of asymptotic amplitudes of the field perturbations. Finally, we provide more details on the background (ultralight DM) solutions considered in our work: self-gravitating DM solitons (either in isolation, or hosting a parasitic black hole), and boson clouds sustained by black holes (BHs).

\section{A crash course on BH perturbation theory}

The natural framework to study EMRIs is BH perturbation theory~\cite{Zerilli:1970wzz, Sago:2002fe, Martel:2005ir}. Consider a secondary compact object of mass $m_p$ sourcing perturbations to a particular (non-vacuum) background solution $\{(\widehat{\mathcal{M}},\widehat{g}_{\mu \nu}),\widehat{\Phi}\}$ of the Einstein-Klein-Gordon (EKG) system, representing a massive BH of mass $M_{\rm BH}$ with a scalar field environment of mass~$M_\Phi$,
\begin{gather}
g_{\mu\nu} = \widehat{g}_{\mu\nu} + q \, \delta g_{\mu\nu} + \mathcal{O}(q^2)\, , \\
\Phi = \widehat{\Phi} + q \, \delta\Phi + \mathcal{O}(q^2) \, .
\end{gather}
Our perturbation scheme is built over the mass-ratio $q\equiv m_p/M_\text{BH}$.

\subsection{Background solution}
We focus on spherically-symmetric backgrounds, which in spherical coordinates have line element
\begin{nofleqn}{equation}
    \dd\widehat{s}^2 \equiv \widehat{g}_{\mu \nu}\dd x^\mu \dd x^\nu\approx -A(r) \mathrm{d}t^2+\frac{\mathrm{d}r^2}{B(r)}+r^2  \mathrm{d}\Omega^2,
    \label{eq:MetricGenProfile}
\end{nofleqn}
where~$\dd\Omega \equiv \dd\theta^2 + \sin^2 \theta \dd\varphi^2$ is the line element on the $2$-sphere, with the background scalar field
\beq
\widehat{\Phi} \approx \phi_0(r) e^{-i \omega t} \, , 
\eeq
where $\omega$ is (in general) a complex-valued parameter found by solving a boundary value problem, imposing regularity at the BH horizon (or at the origin) and at spatial infinity. When taking the backreaction into account, an exact solution of this form with~$\mathrm{Im}\, \omega \neq 0$ cannot possibly exist, since no-hair theorems prevent these spacetimes to be \textit{exactly} static. Nevertheless, the previous expressions can still describe dynamical solutions for an interval of time~$\Delta t \lesssim \tau_{\mathrm{backg}}\equiv(M_{\mathrm{BH}}/M_\Phi) |\mathrm{Im}\, \omega|^{-1}$ (see, e.g., Refs.~\cite{Barranco:2017aes,Cardoso:2022nzc}). Although not needed, for simplicity, in this work we focus on scalar fields that are light enough ($\mu M_{\mathrm{BH}}\ll 1$), so that accretion is suppressed~\cite{Cardoso:2022nzc} and the inspiral happens on a timescale~$\tau_ {\mathrm{insp}}\ll \min \left(\tau_{\mathrm{backg}},|\mathrm{Im}\, \omega|^{-1}\right)$. For such light fields, we can effectively use~$\mathrm{Im}\, \omega \approx 0$ to compute the energy and angular momentum loss rate in EMRIs.

\subsection{Multipolar expansion}
The spherical symmetry of the background solution guarantees that any linear perturbation can be decomposed into irreducible representations of $\text{SO}(3)$. For the scalar field, this corresponds to the standard expansion in (scalar) spherical harmonics~\cite{NIST:DLMF}
\begin{gather}
\delta \Phi = \frac{e^{-i \omega t}}{r}\sum_{\ell,m} \phi_+^{\ell m}\left(t,r\right) Y_{\ell m}(\theta, \varphi) \, , \label{eq:PhiPlus} \\
\delta \Phi^* = \frac{e^{i \omega t}}{r}\sum_{\ell,m} \phi_-^{\ell m}\left(t,r\right) Y_{\ell m}(\theta, \varphi)\, .
\end{gather}
where $\sum_{\ell,m}\equiv\sum_{\ell=0}^{\infty}\sum_{m=-\ell}^\ell$.
We treat the~$\delta \Phi$ and $\delta \Phi^*$ perturbations as independent and factor out the dependence on the background frequency~$\omega$, which as we will see below will allow us to obtain linearized equations of motion whose coefficients do not depend on time. Although in the main text we consider only scalar field structures diluted enough, so that self-interactions are negligible, here for generality we include a U$(1)$-invariant self-interaction potential~$V$. So, it is also useful to expand it, and its first derivative, as
\begin{gather}
V \approx \widehat{V} + \sum_{\ell,m}  \delta V^{\ell m}(t,r) Y_{\ell m}(\theta, \varphi)\, , \\
U \equiv \frac{\partial V}{\partial |\widehat{\Phi}|^2} \approx \widehat{U} + \sum_{\ell,m}  \delta U^{\ell m}(t,r) Y_{\ell m}(\theta, \varphi) \, .
\end{gather}
For diluted scalar field structures, we have simply
\begin{gather}
\widehat{V} \approx \mu^2 \phi_0^2 \, , \qquad \widehat{U} \approx \mu^2 \, ,
 \\
 \delta V^{\ell m} = \frac{\phi_0}{r} \left( \phi_+^{\ell m} + \phi_-^{\ell m} \right)\, , \quad \delta U^{\ell m} \approx 0 \, ,
\end{gather}

The gravitational perturbations can be expanded in the basis of the (ten) tensor spherical harmonics, which can be grouped into polar (electric/even) or axial (magnetic/odd) types, depending on their behavior under parity transformations $\left(\theta, \varphi \right) \to \left(\pi - \theta, \pi + \varphi \right)$~\cite{Sago:2002fe},
\begin{widetext}
\beq
\bm{\delta g}^\text{axial}&=&\sum_{\ell,m}\frac{\sqrt{2\ell(\ell+1)}}{r} \bigg[i\,h_{1}^{\ell m }\bm{c}_{\ell m}
-h_{0}^{\ell m} \bm{c}^{0}_{\ell m} + \frac{\sqrt{(\ell+2)(\ell-1)}}{2}h_{2}^{\ell m }\bm{d}_{\ell m} \bigg]\ , \\
\bm{\delta g}^{\text{polar}}&=&\sum_{\ell,m} A\,H^{\ell m }_0 \bm{a}^{0}_{\ell m}-i\sqrt{2}H^{\ell m}_1 \bm{a}^{1}_{\ell m} +\frac{1}{B}H_2^{\ell m }\bm{a}_{\ell m} +\sqrt{2}K^{\ell m } \bm{g}_{\ell m} \nonumber \\
&&+  \frac{\sqrt{2\ell \left( \ell +1 \right)}}{r}\left(h^{(e) \ell m }_1 \bm{b}^1_{\ell m} - i h^{(e) \ell m }_0 \bm{b}^0_{\ell m} \right) + \left( \sqrt{\frac{(\ell +2 )(\ell +1) \ell (\ell -1 ) }{2} }\bm{f}_{\ell m} - \frac{\ell \left(\ell +1\right)}{\sqrt{2}} \bm{g}_{\ell m} \right) G^{\ell m}   . \label{eq:ExpansionMetric}
\eeq
\end{widetext}
We omit the dependence on $(t,r,\theta,\varphi)$ to avoid cluttering, but the mode perturbations $h_1^{\ell m}, \, h_0^{\ell m}, \dots$ are functions of only $(t,r)$, while $\bm{a}^0_{\ell m}, \, \bm{a}_{\ell m}, \dots$ are the ten tensor spherical harmonics depending only on $(\theta, \varphi)$, which are orthonormal on the $2$-sphere, as defined in Ref.~\cite{Sago:2002fe}.

In a similar fashion, the energy-momentum tensor can be expanded as
\beq
\bm{T}&=&\sum_{\ell,m}
{\cal A}^{0}_{\ell m }\bm{a}^{0}_{\ell m}+{\cal A}^{1}_{\ell m}\bm{a}^{1}_{\ell m}
+{\cal A}_{\ell m }\bm{a}_{\ell m}+{\cal B}^{0}_{\ell m }\bm{b}^{0}_{\ell m}\nonumber \\
&&+ {\cal B}_{\ell m }\bm{b}_{\ell m}\nonumber +{\cal Q}^{0}_{\ell m }\bm{c}^{0}_{\ell m}+{\cal Q}_{\ell m }\bm{c}_{\ell m} +{\cal D}_{\ell m }\bm{d}_{\ell m}\nonumber \\
&&+{\cal F}_{\ell m }\bm{f}_{\ell m}+ {\cal G}_{\ell m}\bm{g}_{\ell m} \ .\label{harmonicexp}
\eeq
For a given source the expansion coefficients can be obtained by projecting the energy-momentum tensor on the respective spherical harmonic, e.g., ${\cal A}^{0}_{\ell m } = (\bm{a}^{0}_{\ell m}, \bm{T})$, with $(\cdot,\cdot)$ the inner product on the $2$-sphere
\beq
\big(\bm{R}^{\ell' m'}, \bm{S}^{\ell m} \big) = \int_{S^2} \dd\Omega  \big(R^{\ell' m'}_{\mu \nu}\big)^*  S^{\ell m}_{\lambda \rho} \, \eta^{\mu\lambda} \,  \eta^{\nu \rho} \, ,
\eeq
where 
\beq
\eta_{\mu\nu} = \text{diag}\left(-1, 1, r^2, r^2 \sin^2 \theta \right) \, .
\eeq
\subsection{Gauge invariance}
The theory of general relativity (GR) is invariant under diffeomorphisms, which translates into the gauge invariance of linear perturbations:
\begin{gather}
x^\mu \to x'^\mu = x^\mu  + q \,\xi^\mu \,, \nonumber \\
\delta g_{\mu\nu} \to \delta g'_{\mu\nu} = \delta g_{\mu\nu} - 2\nabla_{(\mu} \xi_{\nu)} \, , \nonumber \\
\delta \Phi \to \delta \Phi'=\delta \Phi- \xi^\mu \partial_\mu \widehat{\Phi}\,, \nonumber
\end{gather}
where~$\xi^\mu$ is the vector field generating the (infinitesimal) diffeomorphism; we can use its four components to impose some coefficients in the expansion of $\delta g_{\mu\nu}$ to vanish. Note that $\xi^\mu$ can also be expanded in the polar and axial components
\beq
\xi^\mu_\text{polar} &=&  \sum_{\ell,m} \big( - \frac{1}{A}  \xi_t^{\ell m}  ,   B \xi_r^{\ell m} ,   0  ,  0\big)Y_{\ell m} \nonumber \\
&&+ \frac{\xi_\Omega^{\ell m}}{r^2 \sin\theta}  \left(0 , 0 , \sin \theta \,\partial_\theta Y_{\ell m} ,  \partial_\varphi Y_{\ell m} \right) \, , \nonumber\\
\xi^\mu_\text{axial} &=& \sum_{\ell,m} \frac{\xi_\text{ax}^{\ell m}}{r \sin \theta} \left(0 , 0 ,  \partial_\varphi Y_{\ell m} , - \sin \theta\, \partial_\theta Y_{\ell m} \right) \, , \nonumber
\eeq
where the $\xi_t^{\ell m}$, $\xi_r^{\ell m}$, $\xi_\Omega^{\ell m}$, $\xi_\text{ax}^{\ell m}$ are only functions of $(t,r)$. Then, in terms of tensor spherical harmonics, we find
\begin{widetext}
\beq
2 \bm{\nabla} \bm{\xi} &=& \left( 2 \partial_t\xi_t   - A' B \xi_r \right) \bm{a}^{0} - i \sqrt{2} \left( \partial_r \xi_t + \partial_t \xi_r  - \frac{A'}{A} \xi_t \right) \bm{a}^{1} + \left(2 \partial_r \xi_r + \frac{B'}{B} \xi_r \right) \bm{a}\nonumber \\ 
&&- i\frac{\sqrt{2\ell\left(\ell+1\right)}}{r}\left(\xi_t + \partial_t \xi_\Omega\right) \bm{b}^{0} + \frac{\sqrt{2\ell\left(\ell+1\right)}}{r}\left(\partial_r \xi_\Omega + \xi_r - \frac{2}{r}\xi_\Omega \right)\bm{b} + \sqrt{2 (\ell+1)\ell} \, \partial_t \xi_\text{ax} \bm{c}^{0} \nonumber \\
&&- i \sqrt{2 (\ell+1)\ell}  \left( \partial_r \xi_\text{ax} - \frac{\xi_\text{ax}}{r}\right)\bm{c} + i \frac{\sqrt{2\left(\ell +2\right)\left(\ell +1\right)\left(\ell-1\right)}}{r}\xi_\text{ax} \bm{d} + \frac{\sqrt{2\left(\ell+2\right)  \left(\ell+1\right) \ell  \left(\ell-1\right) } }{r^2} \xi_\Omega \bm{f} \nonumber \\
&&+ \frac{\sqrt{2}}{r^2} \left(2rB \xi_r - \ell \left( \ell +1 \right) \xi_\Omega \right)\bm{g}  \, , 
\eeq
\end{widetext}
where the prime denotes a derivative with respect to $r$ and, henceforth, we omit the $(\ell, m)$ indices to avoid cluttering. If we pick $\xi^\mu$ judiciously, we can eliminate one component of the metric perturbations in the axial sector and three in the polar sector. In this work, we adopt the \textit{Regge-Wheeler} gauge, where the gauge is fixed by setting to zero all terms involving angular derivatives of the highest order~\cite{Regge:1957td, Zerilli:1970wzz}
\beq
h_2=h_0^{(e)}=h_1^{(e)}=G=0 \, . \label{eq:RWgauge}
\eeq
    However, for $\ell \leq 1$ this choice does not completely fix the gauge, because some of the tensor spherical harmonics are identically zero: more precisely, $\bm{b}^0 =\bm{b}=\bm{c}^0=\bm{c}=0$ for~$\ell = 0$, and $\bm{d}=\bm{f}=0$ for $\ell \leq 1$. The $\ell \leq 1$ modes do not contribute to the radiative degrees of freedom of the gravitational field, and in vacuum can be removed by a gauge transformation~\cite{Zerilli:1970wzz, Detweiler:2003ci}. However, in the presence of an environment, they carry physical significance through the matter sector and generate dipolar scalar depletion.


\subsection{Point-particle source}

The secondary body is modelled as a point-particle with energy-momentum tensor 
\beq
T_p^{\mu \nu} &=& \frac{m_p }{r^2\sin\theta} \sqrt{\frac{B}{A}}\frac{dt_p}{d\tau}\frac{dx^\mu}{dt}\frac{dx^\nu}{dt} \nonumber \\ &&\times \delta\left[r-r_p (t) \right]\delta\left[\theta - \theta_p (t)  \right]\delta\left[\varphi - \varphi_p(t) \right]\, , \nonumber \\
\eeq
We take the body to be on circular geodesics of the background at a constant radius $r_p$,
\beq
r_p(t) = r_p\,, \quad  \, \theta_p(t) = \frac{\pi}{2}\, ,\quad  \varphi_p(t) = \Omega_p t \, .
\eeq
The orbital angular frequency is $\Omega_p= \sqrt{\smash[b]{A'_p/2r_p}}$, where $A_p \equiv A(r_p)$~\cite{Cardoso:2008bp}. These orbits admit two conserved quantities: the energy and angular momentum per unit rest mass, respectively, given by~\cite{Cardoso:2008bp}
\begin{gather}
\varepsilon_p = A_p \frac{dt_p}{d\tau} = \dfrac{A_p}{\sqrt{A_p - r_p^2 \Omega_p^2}} \, ,   \\ 
l_p = r_p^2 \frac{d\varphi_p} {d\tau} = \frac{r_p^2 \Omega_p }{\sqrt{A_p - r_p^2 \Omega_p^2}} \, .
\end{gather}

For circular orbits, the tensor harmonics expansion~\eqref{harmonicexp} for the energy-momentum tensor of the secondary greatly simplifies. We have ${\cal A}_{\ell m}={\cal A}_{\ell m}^{(1)}={\cal B}_{\ell m}={\cal Q}_{\ell m}=0$, while the non-vanishing coefficients can be expressed in terms of the orbital parameters as follows
\beq
{\cal A}_{\ell m }^{0}&=&\frac{m_p\sqrt{A B}\varepsilon_p}{r^2} Y^\star_{\ell m }\,\delta_{r_p}\ ,\nonumber\\
{\cal B}_{\ell m }^{0}&=&\frac{m_pi\sqrt{AB}l_p}{r^3\sqrt{(n+1)}}\,\delta_{r_p} \,\partial_\phi Y^\star_{\ell m }\ ,\nonumber\\
{\cal Q}_{\ell m }^{0}&=&-\frac{m_p\sqrt{AB}l_p}{r^3\sqrt{(n+1)}}\,\delta_{r_p} \,\partial_\theta Y^\star_{\ell m }\ ,\nonumber\\
{\cal G}_{\ell m }&=&\frac{m_pl_p^2\sqrt{AB}}{r^4\sqrt{2}\varepsilon_p}\,\delta_{r_p} \, Y^\star_{\ell m}\ ,\nonumber\\
{\cal D}_{\ell m }&=&\frac{m_pil_p^2\sqrt{AB}}{\varepsilon_pr^4\sqrt{2n(n+1)}}\,\delta_{r_p} \, \partial_{\theta\phi}Y^\star_{\ell m}
\ ,\nonumber\\
{\cal F}_{\ell m }&=&\frac{m_pl_p^2\sqrt{A B}}{r^42\varepsilon_p\sqrt{2n(n+1)}}\, \delta_{r_p} \, (\partial_{\phi\phi}-\partial_{\theta\theta}) Y^\star_{\ell m}\ , \nonumber \\
\eeq
where 
$n=\ell(\ell+1)/2-1$, $Y^\star_{\ell m}=Y^\star_{\ell m}(\pi/2,0)$
and $\delta_{r_p}=\delta(r-r_p)$. 

\section{Evolution Equations}

We now have all the necessary ingredients to obtain the equations of motion to evolve the perturbations. 

\subsection{$\ell \geq 2$} \label{sec:l2}

We start with the $\ell \geq 2$ modes. We will only focus on the polar sector, where gravitational perturbations couple to matter ones. First, redefine
\beq
H_0 &=& K + \frac{r}{A} S \, , \\
H_1 &=& \frac{r}{A} \tilde{H_1} \, .
\eeq

This choice is inspired in studies of perturbations around relativistic stars~\cite{Allen:1997xj} and the factoring of $r$ and $A$ captures the asymptotic behavior at, respectively, large distances and near the BH horizon. Then, we define 
\beq
\mathcal{E}_{\mu\nu}=G^{(1)}_{\mu\nu}-8\pi(T^{\Phi(1)}_{\mu\nu}+T^p_{\mu\nu})\ ,
\eeq
and follow the same steps as in Ref.~\cite{Cardoso:2022whc}

\begin{itemize}

\item $\mathcal{E}_{\theta\theta} - \mathcal{E}_{\varphi\varphi}/\sin^2 \theta$ gives an algebraic relation for $H_2$
\begin{align}
H_2 &= K +  \frac{r}{A} S - \frac{8 \pi r^2}{\sqrt{n(n+1)/2}} \,  \, \mathcal{F}_{\ell m} \, . 
\end{align}
\item $\mathcal{E}_{r\theta}$ gives a relation for $\partial \tilde{H}_1 / \partial t$ 
\beq
\frac{\partial \tilde{H}_1}{\partial t} &=& \sqrt{\frac{A}{B} } \frac{\partial S}{\partial r_*} + \frac{A}{r}S + \frac{A\,A'}{r}K \nonumber \\
&-& \frac{8\pi\ A^2}{r^2} \phi_0'\left(\phi_+ + \phi_- \right)  \nonumber \\
&-& \frac{ 4\pi \,  A }{\sqrt{n(n+1)/2}}\left(2A + r A' \right)  \mathcal{F}_{\ell m} \, .
\eeq

\begin{widetext}
\item $\mathcal{E}_{tt} - A\, B\, \mathcal{E}_{rr}$ gives the first second-order ``wavelike'' equation for $K$

\begin{align}
&-\frac{\partial^2 K}{\partial^2 t} + \frac{\partial^2 K}{\partial^2 r_*} + 2\frac{\sqrt{A B}}{r}  \frac{\partial K}{\partial r_*} - \left[ \frac{A}{r^2}\left( \ell \left(\ell + 1 \right) + 2B - 2 + 2\,r \, B' \right) - \frac{2 A' B}{r} + 8\pi \omega^2 \phi_0^2 + 8\pi A\, B \,\phi_0'^2 \right] K \nonumber \\
&= \left[ \frac{2B}{r} + 2B' - \frac{2A' B}{A} +  8\pi \frac{r}{A} \omega^2 \phi_0^2 + 8 \pi \,B\, r\, \phi_0'^2 \right]S \nonumber \\
&+ 16\pi \frac{A B}{r^2}\phi_0' \left(\phi_+ + \phi_- \right)   - 8\pi A \, \delta V - 8\pi  \mathcal{A}^0_{\ell m} - \frac{8\pi \, A\, B \, r}{\sqrt{n(n+1)/2}}\frac{\partial \mathcal{F}_{\ell m}}{\partial r} \nonumber \\
& + \frac{4\pi}{\sqrt{n(n+1)/2}}\left[ 2 \, r A' B - A \left(\ell\left(\ell+1 \right) + 2 + 2B + 2 r B' \right) - 8 \pi r^2 \omega^2 \phi_0^2 - 8\pi r^2 A B \, \phi_0'^2 + 8 \pi r^2 A \, \widehat{V} \right] \mathcal{F}_{\ell m}
\end{align}

\item $\mathcal{E}_{\theta\theta} + \mathcal{E}_{\varphi\varphi}/\sin^2 \theta$  gives another second-order ``wavelike'' equation for $S$

\begin{align}
&-\frac{\partial^2 S}{\partial^2 t} + \frac{\partial^2 S}{\partial^2 r_*} - \left[ \frac{A}{r^2}\left( \ell \left(\ell + 1 \right) + 2B - 2 + \frac{r}{2} \, B' \right) + \frac{ A' B}{2r} - 8\pi \omega^2 \phi_0^2 + 8\pi A\, B \,\phi_0'^2 + 8\pi A \, \widehat{V} \right]S\nonumber \\
&= \frac{A}{r} \left[ \frac{2A' B}{r} - A' B' - 2A''B \right]K + \frac{8\pi}{r^2}\left[ \left( A^2 \left(B' - \frac{2B}{r} \right) + 2 A A' B \right) \phi_0' + 2 A^2\, B \, \phi_0'' \right]\left( \phi_+ + \phi_- \right)  \nonumber \\
&+8\pi \sqrt{2} \, \frac{A^2}{r} \mathcal{G}_{\ell m} - \frac{8\pi A\, r}{\sqrt{n(n+1)/2}}\frac{\partial^2 \mathcal{F}_{\ell m}}{\partial t^2} + \frac{4\pi\, A\, B}{\sqrt{n(n+1)/2}}\left(2A + rA' \right) \frac{\partial \mathcal{F}_{\ell m}}{\partial r} \nonumber \\
& + \frac{4\pi}{\sqrt{n(n+1)/2}}\frac{A}{r}\Bigg[  2 r A' B + r^2 B \frac{A'^2}{A} - A \left( \ell \left(\ell + 1 \right) - 4 + 4B  \right) + 16\pi r^2 \omega^2  \phi_0^2 -16\pi r^2 A B \, \phi_0'^2 - 16 \pi r^2 A \widehat{V}  \Bigg] \mathcal{F}_{\ell m}
\end{align}

\item Finally, the Klein-Gordon  perturbed up to first-order gives
\begin{align}
&-\frac{\partial^2 \phi_+}{\partial^2 t} + \frac{\partial^2 \phi_+}{\partial^2 r_*} + 2 i \, \omega \frac{\partial \phi_+}{\partial t} + \left[\omega^2 - \frac{A}{r^2}\ell(\ell +1) - \frac{A B' + A' B}{2r} -  A \widehat{U}  \right]\phi_+ -8 \pi A B\, \phi_0'^2 \, \left(\phi_+ + \phi_- \right) -  r\, A \, \phi_0 \,  \delta U \nonumber \\
&= -2i r \omega \phi_0 \frac{\partial K}{\partial t} - i \frac{r^2}{A} \omega \phi_0 \frac{\partial S}{\partial t} + i \frac{r^2}{A} \sqrt{\frac{B}{A}} \omega \phi_0 \frac{\partial \tilde{H}_1}{\partial r_*}  - \left[\frac{r^2}{A}\omega^2\phi_0 + r B \phi'_0 \left(r\frac{A'}{2A} - r\frac{B'}{2B} -2\right)  - r^2 B \phi_0'' \right]\left(S + \frac{A}{r} K \right) \nonumber  \\
&+ i \frac{r^2 \omega}{2A}\left[\left(6 \frac{B}{r} + B' - 3 \frac{A' B}{A}  \right) \phi_0 + 4B\phi'_0 \right]H_1 \nonumber \\
&+ i \frac{4\pi r^3 \omega \phi_0}{\sqrt{n(n+1)/2}}\frac{\partial \mathcal{F}_{\ell m}} {\partial t} - 	\frac{4\pi r^3 A B \, \phi'_0}{\sqrt{n(n+1)/2}}\frac{\partial \mathcal{F}_{\ell m}} {\partial r} - \frac{4\pi \, r^2 A B}{\sqrt{n(n+1)/2}}\left[ \left(4 + r \frac{B'}{B} \right) \phi_0'	 + 2 r \phi_0'' \right ] \mathcal{F}_{\ell m} \, ,
\end{align}
%
\end{widetext}	
The equation for $\phi_-$ is similar but with the complex conjugate coefficients multiplying the perturbations.

\end{itemize}

\subsection{$\ell = 1$} \label{sec:l1}

As mentioned above, for $\ell = 1$, our gauge choice in Eq.~\eqref{eq:RWgauge} is not completely fixed because some of the spherical harmonics are identically 0. In particular, $\bm{g}=0$ so we can fix one more gravitational perturbation function to $0$. We follow the original choice of Zerilli~\cite{Zerilli:1970wzz} and pick $
K = 0 $. Then

\begin{widetext}

\begin{itemize}

\item $\mathcal{E}_{tr}$ yields an algebraic relation for $H_1$

\beq
H_1 &=& -\frac{1}{AB + r AB' + 4\pi r^2 \left( A \widehat{V} - \omega^2 \phi_0^2 + AB\phi_0'^2 \right) } \Bigg[ r A \frac{\partial H_2}{\partial t} + i 4\pi \omega A \left( \phi_0 + r \phi_0' \right)  \left(\phi_+ - \phi_-\right) \nonumber \\
&-&  i 4\pi  \omega \sqrt{\frac{A}{B}} r \phi_0 \left(\frac{\partial \phi_+}{\partial r_*} - \frac{\partial \phi_-}{\partial r_*}\right) - 4\pi A r \phi_0' \left( \frac{\partial \phi_+}{\partial t} +  \frac{\partial \phi_-}{\partial t}\right) \, .
\Bigg]
\eeq

\item $\mathcal{E}_{r\theta}$ gives a relation between $\partial H_0/\partial r_*$ and the other variables which can then be substituted in $\mathcal{E}_{rr}$ to obtain an algebraic relation for $H_0$

\beq
H_0 &=& \frac{1}{2A\left(B -1\right) - r A' B+ 8 \pi r^2 \omega^2 \phi_0^2} \Bigg[ 2 r B \frac{\partial H_1}{\partial t} + \Big(2A\left(B-1\right) + r A' B + 8 \pi r^2 \widehat{V} -  8\pi r^2 \omega^2 \phi_0 \Big) H_0 \nonumber \\
&-&8\pi\left[ \left( r \omega^2 \phi_0 + AB \phi_0' \right) \left( \phi_+ + \phi_- \right) + r \sqrt{A B} \phi_0'\left( \frac{\partial \phi_+}{\partial r_*} + \frac{\partial \phi_-}{\partial r_*} \right) +  i r \omega \phi_0 \left( \frac{\partial \phi_+}{\partial t} + \frac{\partial \phi_-}{\partial t} \right)  \right]\nonumber \\
\eeq

The two equations above can then be substituted in $\mathcal{E}_{tt}$ to obtain an equation for $\partial H_2 /\partial r_*$
\beq
\frac{\partial H_2}{\partial r_*} &=& -\frac{1}{r}\sqrt{\frac{A}{B}}\Bigg[ \left(1 + B + rB' + 4\pi r^2 B \phi_0'^2\right)H_2 + \left( B - 1 + r B'+ 4\pi r^2 \left(\widehat{V} + B \phi_0' \right)\right) H_0 \Bigg] \nonumber \\
&+& \frac{4\pi}{r \sqrt{A B}} \left[ \left( r \omega^2 \phi_0 - AB \phi_0' \right)\left( \phi_+ + \phi_- \right) + i  r \omega \phi_0 \left( \frac{\partial \phi_+}{\partial t} + \frac{\partial \phi_-}{\partial t} \right) +  \sqrt{A B}  r \phi_0' \left( \frac{\partial \phi_+}{\partial r_*} + \frac{\partial \phi_-}{\partial r_*} \right) \right] \nonumber  \\
&+& \frac{8\pi r }{\sqrt{AB}} \mathcal{A}_0 \, ,  \label{eq:H2r}
\eeq

\item Finally, the perturbed Klein-Gordon equation in this gauge choice is written as
\begin{align}
&-\frac{\partial^2 \phi_+}{\partial t^2 } + \frac{\partial^2 \phi_+}{\partial^2 r_*} + 2 i \, \omega \frac{\partial \phi_+}{\partial t} + \left[\omega^2 - \frac{A}{r^2}\ell(\ell +1) - \frac{A B' + A' B}{2r} -  A \widehat{U}  \right]\phi_+ - r A \delta U \phi_+ \nonumber \\
&= \frac{r}{2} \sqrt{A B}\phi_0' \frac{\partial H_0}{\partial r_*} + i r \omega \sqrt{\frac{B}{A}} \phi_0 \frac{\partial H_1}{\partial r_*} + \frac{r}{2} \sqrt{AB} \phi_0' \frac{\partial H_2}{\partial r_*} - i\frac{r}{2} \omega \phi_0 \frac{\partial H_0}{\partial t} - r B \phi_0' \frac{\partial H_0}{\partial t} - i \frac{r}{2} \omega \phi_0 \frac{\partial H_2}{\partial t} \nonumber \\
&- r \omega^2 \phi_0 H_0 + i \omega \left[ \phi_0 \left(2B + \frac{r}{A}\left(AB' - A'B\right) \right) +2 B r \phi_0' \right]H_1
\nonumber + \left[ \left(2 AB + \frac{r}{2} \left( AB' + A' +B\right) \right) \phi_0' + r A B \phi_0''\right]H_2 \, , \\
\end{align}
and the same equation for $\phi_-$ but with complex conjugated coefficients.

\end{itemize}

\end{widetext}

\section{Background solutions}

In this section, we provide more details on the different backgrounds used for the metric and the scalar field distribution. First, let us rewrite the $g_{rr}$ metric function in terms of the Misner-Sharp mass function
\beq
B(r) = 1-\frac{2m(r)}{r} \, .
\eeq
The EKG system at  $0$-th order can then be written as~\cite{Barranco:2017aes} 
\begin{widetext}   
\beq
m' &=& 2\pi r^2 \left(1 - \frac{2m}{r} \right)\left( \left| \phi_0' \right|^2 
 + \frac{\mu^2 + \frac{\omega^2}{A} }{1-\frac{2m}{r}}  \left| \phi_0 \right|^2 \right)\, ,  \label{eq:MassODE}\\
\log'(A) &=& \frac{2m}{r^2 \left(1 -\frac{2m}{r} \right)}  + 2\pi r \left( \left| \phi_0' \right|^2
 - \frac{\mu^2 - \frac{\omega^2}{A} }{1-\frac{2m}{r}}  \left| \phi_0 \right|^2  \right) \, , \label{eq:AODE} \\
\left(1-\frac{2m}{r}\right) \phi_0'' &+&\sqrt{\frac{1-\frac{2m}{r}}{A}}\frac{1}{r^2}\left(r^2 \sqrt{A \left(1-\frac{2m}{r}\right)}  \right)' \phi_0'= \left(\mu^2 - \frac{\omega^2}{A} \right) \phi_0 \, . \label{eq:PhiODE}
\eeq

\end{widetext}

\subsection{Newtonian Boson Stars}

\textit{Newtonian} Boson Stars (NBS) are localized, low-compactness, self-gravitating solutions of the system above, where the scalar field is necessarily weak, i.e. $\left|\phi_{0, \rm{NBS}}(r) \right| \ll 1$. When this is verified, the background metric function is approximately~\cite{Annulli:2020lyc} 
\beq
A(r) \approx 1 + 2 U_{\rm NBS}(r) \, , 
\eeq
where $\left|U_{\rm NBS}(r)\right| \ll 1$ is the Newtonian gravitational potential of the NBS. The frequency of the fundamental mode is given by
\beq
\omega = \mu - \gamma \, , 
\eeq
where $0 < \gamma \ll \mu $. As we will see below, Eqs.~\eqref{eq:MassODE}-\eqref{eq:PhiODE} can be reduced to the Schrödinger-Poisson system for $\phi_{0, \rm NBS}$ and $U_{\rm NBS}$, respectively. This forms an eigenvalue problem for $\gamma$ where one needs to impose regularity of $\phi_{0, \rm NBS}$ and $U_{\rm NBS}$ at the origin and demand asymptotic flatness at large distances. It is possible to show that for large $r$, the scalar field falls off according to a Yukawa potential $\phi_{0, \rm NBS} \sim e^{-\sqrt{2\mu \gamma}r} /r$,  while the gravitational potential behaves as $ \left| U_{\rm NBS} \right| \sim M_\Phi/r$, where $ M_\Phi$ is the total mass of the scalar configuration
\beq
M_\Phi = 4\pi \mu^2 \int_0^\infty r'^2 \left| \phi_{0}(r') \right|^2 \, .
\eeq

Ref.~\cite{Annulli:2020lyc} obtained semi-analytic descriptions of the field and gravitational potential accurate up to $1\, \%$ with respect to numerical solutions. We reproduce it here for completeness
\begin{gather}
U_{\rm NBS} = \mu^2 M_\Phi^2\,f(x) \, , \\
f=\frac{a_0+11\frac{a_0}{r_1}x+\sum_{i=2}^{9}a_ix^i-x^{10}}{(x+r_1)^{11}}\,,\nonumber\\
x=\mu^2M_{\Phi}r\,,r_1=1.288\,,\nonumber\\
a_0=-5.132\,,a_2=-143.279\,,a_3=-645.326\,,\nonumber\\
a_4=277.921\,, a_5=-2024.838\,,a_6=476.702\,,\nonumber\\
a_7=-549.051,\, a_8=-90.244\,,a_9=-13.734\,.\nonumber
\end{gather}
and
\begin{gather}
\phi_0 = \mu^2 M_\Phi^2 g(x) \, , \\
g=e^{-0.570459 x}\frac{\sum_{i=0}^{8}b_ix^i+b_fx^{9.6}}{(x+r_2)^{9}}\,,\nonumber\\
x=\mu^2M_{\Phi}r\,,r_2=1.182\,,\nonumber\\
b_0=0.298\,,b_1=2.368\,,b_2=10.095\,,\nonumber\\
b_3=12.552\,, b_4=51.469\,,b_5=-8.416\,,\nonumber\\
b_6=54.141,\, b_7=-6.167\,,b_8=8.089\,,\nonumber\\
b_f=0.310\,.
\end{gather}
and the frequency of the fundamental mode is
\beq
\gamma \approx 0.163 M^2_\Phi \mu^3 \, .
\eeq

As mentioned in the main text, NBSs exhibit a scale invariant mass-radius relation
\beq
M_\Phi	\approx 9 \times  10^{8}M_\odot \left(\frac{0.1\, \text{kpc}}{R_\text{98}}\right) m_{22}^{-2} ,
\eeq
where we recall~$m_{22}\equiv m_\Phi/10^{-22}\,\mathrm{eV}$,~$R_{98}$ is the radius enclosing $98\%$ of~$M_\Phi$.

An important property of NBSs is that near their center, in particular for radius $r \lesssim 0.1 R_{98}$, both the scalar field profile and the gravitational potential are approximately constant
\beq
\phi_{0, \rm{NBS}}(0) &\approx& 0.0662 M_\Phi^2 \mu^3\, , \\
U_{\rm NBS} (0) &\approx& - \frac{1}{2}\frac{M_\Phi}{R_{98}} \, .
\eeq
All the features discussed are illustrated in Fig.~1 of Ref.~\cite{Annulli:2020lyc}.

\subsection{Parasitic Black Holes}

Now we want to place a BH in the center of a NBS. As discussed in the main text, no-hair theorems~\cite{Ruffini:1971bza,Bekenstein:1972ny} prevent the existence of a static, spherically-symmetric solution in this setup. Consequently, the scalar field distribution will be accreted by the BH and $\text{Im}(\omega) \neq 0$. However, Numerical Relativity results supported by analytic estimates have shown that when the mass of the NBS is much larger than that of the BH, i.e., $M_{\rm BH} \ll M_{\Phi}$, and the star is Newtonian, i.e., $ \mu M_\Phi \ll 1$, the accretion timescale is larger than the Hubble time~\cite{Cardoso:2022nzc}. In particular~\cite{Cardoso:2022nzc}
\beq
\text{Im}(\omega) \sim 10^{-1} \mu^3 M_\Phi^2 \, ,
\eeq
and the time $\tau_{10\%}$ necessary for the BH to accrete $10\%$ of the initial mass of the NBS, $M_{\Phi, 0}$, is~\cite{Cardoso:2022nzc}
\beq
\frac{\tau_{10\%}}{10\, \text{Gyr}} \lesssim 6.6 \left(\frac{10^{10}}{M_{\rm \Phi, 0}} \right)^5 \left( \frac{10^{-22} \, \text{eV}} {\mu}\right)^6 \, .
\eeq
Therefore, for the inspiral timescale of an EMRI we can take $\text{Im}(\omega) \approx 0$. 

The hierarchy of length scales determined by the event horizon and the NBS's radius allows us to separate the problem into two distinct regions. Close to the event horizon, the gravitational field is dominated by BH, while at far-away distances, gravity is dictated by the self-gravity of the NBS. In between, there is a transition radius where the two gravitational fields are comparable. It is useful to define the influence radius $r_i$ as the location where they match
\beq
\frac{M_{\rm BH}}{r_i} &=& \left|U_{\rm NBS}(r_i) \right| \Leftrightarrow \nonumber \\
\Leftrightarrow  r_i &=& \frac{M_{\rm BH}}{\left|U_{\rm NBS}(r_i) \right|} \sim 2 \frac{M_{\rm BH}}{M_\Phi} R_{98} \, .
\eeq

Since $M_{\rm BH} \ll M_{ \rm NBS}$ and the star is Newtonian, so $R_{\rm NBS} \gg M_\Phi$, the transition radius satisfies $2M_{\rm BH} \ll r_i \ll R_{\rm{NBS}}$, meaning it is located far-away from the BH horizon but still deep inside the NBS. 

\subsubsection{Inner region ($r \ll r_i$)}

In this region, we take the test-field approximation and consider that the scalar field does not back react on the spacetime metric. The ordinary differential equation (ODE) for the mass function, Eq.~\eqref{eq:MassODE}, becomes simply 
\beq
m'(r) = 0 \Leftrightarrow m(r) = \text{const} = M_{ \rm BH } \, , \label{eq:InnerSolMass} 
\eeq
where we impose that at the BH horizon this constant is the BH mass. 

Having this, Eq.~\eqref{eq:AODE} reduces to
\beq
\log'(A) &=& \frac{2M_{\rm BH}}{r^2\left(1-\frac{2M_{\rm BH}}{r}\right)} \Leftrightarrow \nonumber \\
\Leftrightarrow \log(A) &=&  \log\left(1-\frac{2M_{\rm BH}}{r}\right) + \Upsilon \Leftrightarrow \nonumber \\
\Leftrightarrow A(r) &=& \left(1-\frac{2M_{\rm BH}}{r}\right) e^\Upsilon \, , \label{eq:InnerSolA}
\eeq
where $\Upsilon$ is a constant which will be fixed by demanding continuity of $A$ with the outer region, where the NBS self-gravity dominates over the BH potential. 

Similarly, the Klein-Gordon equation is solved by~\cite{Annulli:2020lyc}
\beq
\phi_0(r) = \mathcal{C} \left(1-\frac{2M_{\rm BH}}{r}\right)e^{-2i\mu M_{\rm BH}}
\eeq
where $\mathcal{C}$ is a constant, which will also be fixed by demanding continuity of $\phi_0$ with the outer region.

\subsubsection{Outer region ($r \gg r_i$)}

We now focus on the region at large radius, far-away from the radius of influence of the BH. Here, the scalar field is non-relativistic and the radial derivatives of $\phi_0$ are subleading in Eqs.~\eqref{eq:MassODE} and~\eqref{eq:AODE}. Then, let us take the following ansatz for $A$ and $\phi_0$
\beq
A &=& \left(1 - \frac{2M_{\rm BH}}{r} \right) e^{2 U(r)} \, , \label{eq:GttGenericTerm}\\
\phi_0 &=& \left(1 - \frac{2M_{\rm BH}}{r} \right)^{-2i\mu M_{\rm BH}} \tilde{\phi}_0(r) \, , 
\eeq
where $U(r)\sim \mathcal{O}\left( \left| \phi_0 \right|^2 \right)$ is a function to be determined, which satisfies $\left|U(r) \right| \gg M_{\rm BH}/r$ in this region. Plugging these in Eqs.~\eqref{eq:MassODE}-\eqref{eq:PhiODE} and keeping only the leading terms in $1/r$ and $\left| \phi_0 \right|^2$
\beq
m(r) &=& M_{\rm BH} +  4\pi \mu^2 \int_{r_i}^{r} dr' r'^2 \left| \phi_0(r') \right|^2 \, ,  \\
U' &=& \frac{4\pi \mu^2}{r^2} \int_{r_i}^{r} dr' r'^2 \left| \phi_0(r') \right|^2  \, , \\
\tilde{\phi}_0'' &+& \frac{2}{r}\tilde{\phi}_0' - 2\mu \left(\mu U + \gamma \right) \tilde{\phi}_0 = 0 \, . \label{eq:SchrodingerDerived}
\eeq
Here, we directly integrated the ODE for $m(r)$ and imposed continuity with respect to mass function in the inner region obtained above (Eq.~\eqref{eq:InnerSolMass}). 

Differentiating the ODE for $U$ with respect to $r$
\beq
U'' &+& \frac{8\pi \mu^2}{r^3} \int_{r_i}^{r} dr' r'^2 \left| \phi_0(r') \right|^2  = \frac{4\pi \mu^2}{r^2} \left|\phi_0 \right|^2 \Leftrightarrow \nonumber \\
\Leftrightarrow U'' &+& \frac{2}{r} U' =  \frac{4\pi \mu^2}{r^2} \left|\phi_0 \right|^2 \, , 
\eeq
which we identify as the Poisson equation for the gravitational potential $U$ being sourced by the scalar field distribution. Therefore, we identify $U$ with the NBS potential $U_{\rm NBS}$ described in the previous section. Since $2M_{\rm BH} \ll r_i \ll R_{\rm NBS}$, for $r \sim r_i$ the potential $U_{\rm NBS}$ is already approximately constant. Therefore, demanding continuity of $A(r)$ between the inner and outer region,  we can match the constant $\Upsilon$ in the inner solution~\eqref{eq:InnerSolA} with the central value of the NBS's gravitational potential, arriving at $\Upsilon = 2 U_{\rm NBS}(0)$.

Finally, we identify the Eq.~\eqref{eq:SchrodingerDerived} as the (time-independent) Schrödinger equation for $\tilde{\phi}_0$, meaning it admits the same profile as the NBS of the previous section. Thus, similarly to the gravitational potential, in the matching region $\phi_0$ is approximately constant and we can match the constant $\mathcal{C}$ of the inner solution~ with the central value of the scalar field, $\mathcal{C}=\phi_0(0)$.

\subsection{Generic background for diluted configurations of matter around BHs}

The background spacetime obtained in the previous section for the case of parasitic BHs has similar properties to other diluted distributions of matter surrounding BHs. In particular, both for anisotropic fluids~\cite{Cardoso:2022whc} and superradiant (scalar) clouds~\cite{Brito:2023pyl}, the $g_{tt}$ metric function reduces to the Schwarzschild one with an exponential redshift pre-factor, as in Eq.~\eqref{eq:GttGenericTerm}. 

We now show this is a generic property when the compactness of the matter configuration is much smaller than the BH compactness. The $tt$ and $rr-$component of Einstein's equations given, respectively, in Eq.~\eqref{eq:MassODE} and Eq.~\eqref{eq:AODE} can be written for an arbitrary (spherically-symmetric) energy-momentum tensor as~\cite{Macedo:2013jja}
\beq
m'&=& 4\pi r^2 \rho  \, , \label{eq:EqMassFluid}\\
\mathrm{log}'(A)&=& \frac{2m}{r^2\left(1 - \frac{2m}{r} \right)} + \frac{4\pi r}{1 - \frac{2m}{r} } \, p_\mathrm{rad} \, , \label{eq:EqAFluid}
\eeq
where 
\beq
\rho = - T_t^t\ \quad , \quad p_\mathrm{rad} = T_r^r \, , 
\eeq
are, respectively, a matter density and radial pressure term. For the case of scalar fields we identify 
\beq
\rho_\phi &=&  \frac{1}{2}\left(1 - \frac{2m}{r} \right)\left( \left| \phi_0' \right|^2 
 + \frac{\mu^2 + \frac{\omega^2}{A} }{1-\frac{2m}{r}}  \left| \phi_0 \right|^2 \right) \, , \nonumber \\ \\
p_\mathrm{rad}^\phi &=& \frac{1}{2}\left(1 - \frac{2m}{r} \right)^{-1}\left( \left| \phi_0' \right|^2
 - \frac{\mu^2 - \frac{\omega^2}{A} }{1-\frac{2m}{r}}  \left| \phi_0 \right|^2  \right)\, . \nonumber \\ 
\eeq

We now repeat the same steps as for the parasitic BH. In the inner region, we consider the matter density to be negligible, from which we directly obtain Eq.~\eqref{eq:InnerSolMass} and~\eqref{eq:InnerSolA}, and the BH solely controls the dynamics.

In the outer region, we impose an ansatz for $A$ as in Eq.~\eqref{eq:GttGenericTerm} and substituting in Eq.~\eqref{eq:EqAFluid} and expanding to first order in $1/r$
\beq
m(r) &=& M_{\rm BH} +  4\pi \int_{r_i}^{r} dr' r'^2 \rho (r') \, , \\
U' &=& \frac{4\pi}{r^2} \int_{r_i}^{r} dr' r'^2 \rho (r') + 4\pi r \, p_\mathrm{rad}  \, ,
\eeq
where again imposed continuity with the mass function. 
For Newtonian configurations, the pressure term is negligible with respect to the matter density one. Differentiating with respect to $r$ we arrive at
\beq
U'' + \frac{2}{r}U' = \frac{4\pi}{r^2} \rho \, , 
\eeq
which once more is Poisson's equation for the Newtonian potential $U$ of the generic matter distribtuion.  


\section{Flux formulas and additional results}

As explained in the main text, the energy fluxes can be written in terms of the perturbations through their respective stress-energy tensors. In the case of gravitational perturbations, we use the Isaacson stress-energy tensor~\cite{Isaacson:1968zza,Martel:2005ir}. The expressions are identical to the Schwarzschild case, as asymptotically the spacetimes are the same (with their respective masses). As such, we can just construct the Zerilli-Moncrief $\psi_Z$ function through $H_1$ and $K$~\cite{Sago:2002fe} and compute the flux through
\begin{equation}
   ^{\ell} \dot{E}^g_{\infty}=\frac{1}{64\pi}\frac{(\ell+2)!}{(\ell-2)!}|\dot{\psi}_{Z}|^2,~~\ell+m~~{\rm even},
\end{equation}
where the explicit form for $\psi_Z$ can be seen in Ref.~\cite{Sago:2002fe}. Alternatively, we can also extract the amplitude through the asymptotic form of $K$, as
\begin{equation}
    K(r\to\infty)\approx\frac{\partial \psi_Z}{\partial r},
\end{equation}
and use this to compute the flux. For the scalar fluxes, we use the stress-energy tensor for the scalar field. The expression for this case can be read directly from Ref.~\cite{Annulli:2020lyc,Brito:2023pyl}, being
\begin{align}
    ^{\ell}\dot{E}^\Phi_{\infty}=&(\omega+\ell\Omega_p){\rm Re}\left[\sqrt{(\omega+\ell\Omega_p)^2-\mu^2}\right]\times\nonumber\\
    &|\phi_+(r\to\infty)|^2,
\end{align}
where $\phi_+$ above is the Fourier amplitude of the scalar field perturbation defined in Eq.~\eqref{eq:PhiPlus}.

\begin{figure}
    \centering
    \includegraphics[width=\columnwidth]{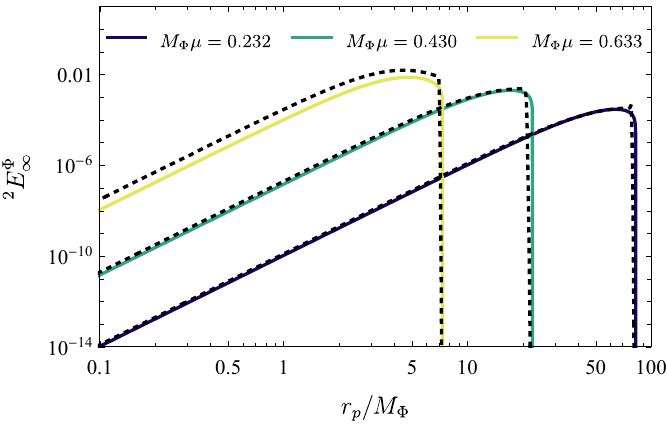}
    \caption{Comparison between the NBS analytical expression (solid lines) with the full relativistic BS fluxes (dotted lines). In the regime of validity of the Newtonian approximation, the analytical expression matches very well our numerical the results. Even in more compact configurations, the analytical expression reproduces the correct behavior of the fluxes with the orbital radius.}
    \label{fig:comp_BS}
\end{figure}

We tested the above numerical relativistic expressions for BS by comparing the scalar fluxes with the analytical NBS presented in~\cite{Annulli:2020lyc} (Eq.~($149$) there), accounting for relativistic modifications regarding the scalar field amplitude at the particle position and angular frequency 
\begin{equation}
 \phi_+(r\to\infty)\approx -\frac{8 \pi ^{\frac{3}{2}} m_p\, \phi_0(r_p) m^{\frac{m}{2}-1} Y_m^m\left(\frac{\pi }{2},0\right) \left(r_p \Omega\right)^{m}}{2^{\frac{m}{2}+2} (\Omega/\mu)^{\frac{m}{2}+1} \Gamma \left(m+\frac{3}{2}\right)}.
\end{equation}
The result can be seen in Fig.~\ref{fig:comp_BS}. The relativistic numerical results match the analytical one in the regime of validity predicted in the Newtonian approximation. Even going beyond it, the analytical prediction works remarkably well to reproduce the functional behavior of the flux.

\begin{figure}
    \centering
    \includegraphics[width=\columnwidth]{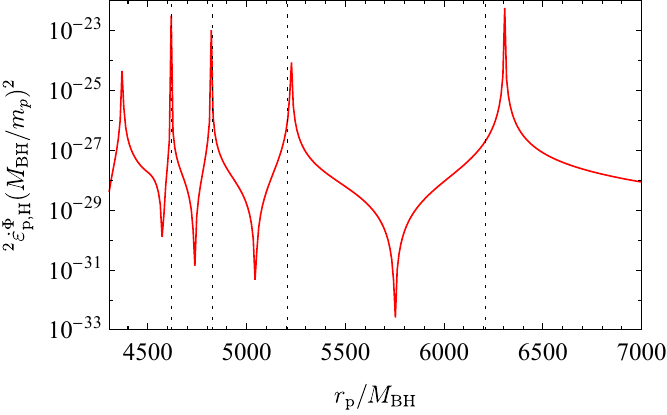}
    \caption{Energy loss rate due to the quadrupolar scalar flux at the horizon for an EMRI evolving in the same NBS with $M_{\Phi}\mu = 0.23$ hosting a parasitic BH of mass $M_{\rm BH} / M_{\Phi} = 0.02$ as in Fig.~1 of the main text. The peaks correspond to the resonant excitation of the proper modes of the scalar configuration, which produces a resonant accretion of scalar by the parasitic BH.}
    \label{fig:flux_hor}
\end{figure}

In the main text, we mention the energy flux going through the horizon in the resonant regime in the case of parasitic BHs. The same procedure used above can be transplanted to compute the fluxes going into the BH. With the amplitude of the scalar field at $\phi_+(r_H)$, the scalar flux can be obtained through
\begin{equation}
    ^{\ell}\dot{E}^\Phi_{r_H}=(\omega+\ell \Omega_p)^2|\phi_+(r_H)|^2.
\end{equation}
With the above expression, we can compute the rate $^\ell E^\Phi_{\rm p,H}$ explained in the main text. 
In Fig.~\ref{fig:flux_hor} we show the quadrupole rate flux into the horizon as a function of the orbital radius. It is evident that for this configuration it is orders of magnitude smaller than the rate of change through the fluxes at infinity (see Fig.~1  in the main text). Nonetheless, it is interesting to see the resonant behavior of the fluxes going into the horizon, indicating that resonant configurations can enhance scalar accretion onto the central BH.

\bibliography{References}